\documentclass[aps,prd,twocolumn,showpacs,amsmath,amssymb]{revtex4-1}
\usepackage{amsmath}
\usepackage{graphicx}
\usepackage{subfigure}
\usepackage{epstopdf}
\usepackage{color}
\usepackage{multirow}
\usepackage{setspace}
\usepackage{overpic}
\usepackage{amssymb}
\usepackage{braket}
\usepackage{makecell}
\usepackage[bookmarksnumbered, pdfstartview=FitH,colorlinks,urlcolor=blue, citecolor=blue,linkcolor=blue] {hyperref}
\usepackage{lineno}
\usepackage{bm}
\usepackage{rotating}
\usepackage[utf8]{inputenc}
\usepackage{orcidlink}

\let\oldequation\equation
\let\oldendequation\endequation
\renewenvironment{equation}
 {\linenomathNonumbers\oldequation}
 {\oldendequation\endlinenomath}

\begin{document}
\title{\boldmath Study of the up-down asymmetry in $B^+\to K^+\pi^+\pi^-\gamma$ \\ using $K_1(1270)$ amplitudes constrained by semileptonic $D$ Decays}

\author{Yijia Zeng\,\orcidlink{0009-0005-3279-0304}}
\affiliation{Institute of High Energy Physics,
	Chinese Academy of Sciences, Beijing 100049, China}

\author{Hailong Ma\,\orcidlink{0000-0001-9771-2802}}
\affiliation{Institute of High Energy Physics,
	Chinese Academy of Sciences, Beijing 100049, China}

\begin{abstract}

Using $K_1(1270)\to K\pi\pi$ amplitude models constrained by the charmed-meson semileptonic decays $D^{+(0)}\to K^-\pi^+\pi^{0(-)} e^+\nu_e$, we present a model-dependent estimate of the up-down asymmetry observed in $B^+\to K^+\pi^+\pi^-\gamma$ in the mass region $M_{K^+\pi^+\pi^-}\in[1.1,1.3]$~GeV/$c^2$. Estimates based on an isolated $K_1(1270)$ show significant
model dependence, but systematically lie below the measured
asymmetry $\bar{\mathcal{H}}_{\mathrm{ud}}=(0.092\pm0.023)$ when assuming
$\lambda_\gamma=1$. We also explore interference with the $K_1(1400)$ tail and find that it can appreciably reduce the tension.

\end{abstract}

\maketitle

\section{Introduction}

In the Standard Model (SM), the left-handed coupling of the $W$ boson to quarks makes the photon in $b\to s\gamma$ transitions predominantly left-handed~\cite{LHCb:2014vnw}, with the right-handed component suppressed by a factor of order $m_s/m_b$. On the other hand, several extensions of the SM~\cite{Atwood:1997zr,Becirevic:2012dx,Paul:2016urs} predict that the photon acquires a significant right-handed component through beyond-the-SM couplings. Measuring the photon polarization in $b\to s\gamma$ therefore provides a sensitive test of the SM and possible new physics.

LHCb has presented an observation of photon polarization in the $b\to s\gamma$ transition through the up-down asymmetry~\cite{Wang:2019wee}
\begin{equation}
\label{eq:Aud_definition}
\mathcal{A}_{ud}=\lambda_{\gamma}\times \frac{3}{4}\times{{\rm Im}[\hat{n}\cdot(\vec{J}\times\vec{J}^*)]}/{|\vec{J}|^2}
\end{equation}
in the $B^+\to K^+\pi^+\pi^-\gamma$ decay~\cite{LHCb:2014vnw}. In particular, a nonzero $\bar{\mathcal{A}}_{ud}=(0.069\pm0.017)$ is observed in the mass region $M_{K^+\pi^+\pi^-}\in[1.1,1.3]$~GeV/$c^2$ dominated by $K_1(1270)$. However, one cannot directly extract photon polarization from $\mathcal{A}_{ud}$ because of the additional hadronic factor $\mathcal{H}_{ud}\equiv{{\rm Im}[\vec{n}\cdot(\vec{J}\times\vec{J}^*)]}/{|\vec{J}|^2}$, where $\vec{J}$ describes the $K_1(1270)\to K\pi\pi$ decay amplitude with $\mathcal{A}(K_1\pi\pi)=\vec{\epsilon}_{K_1}\cdot \vec{J}$.

Determining the hadronic factor $\mathcal{H}_{\mathrm{ud}}$ requires detailed knowledge of the $K_1(1270)\to K\pi\pi$ hadronic current, including the magnitudes and relative phases of its intermediate amplitudes. Previous phenomenological studies have estimated this factor for $K_1(1270)$ and $K_1(1400)$ using amplitude models and experimental inputs~\cite{Gronau:2001ng,Gronau:2002rz,Kou:2010kn,Tayduganov:2011ui,Gronau:2017kyq}. However, their predictions are sensitive to the assumed phenomenological and amplitude models. To reduce this dependence, Refs.~\cite{Bian:2021gwf,Wang:2019wee,Kou:2016iau}
proposed extracting the relevant hadronic information from angular analyses of the $D\to\bar{K}_1(1270)e^+\nu_e$ and $B\to J/\psi K_1(1270)$ decays without specifying the detailed $K_1(1270)\to K\pi\pi$ substructure. However, with only angular distributions, the model-independent approach requires much larger data samples to obtain an effective constraint.

In 2025, BESIII reported a model-independent extraction of $\mathcal{H}_{\mathrm{ud}}=(0.01\pm0.11)$ in $D\to\bar{K}_1(1270)e^+\nu_e$~\cite{BESIII:2025hdt} via a two-dimensional angular fit following Refs.~\cite{Bian:2021gwf,Wang:2019wee}. This result is consistent with the value $\bar{\mathcal{H}}_{\mathrm{ud}}=4\bar{\mathcal{A}}_{\mathrm{ud}}/(3\lambda_\gamma)=(0.092\pm0.023)$
inferred from the LHCb measurement under the assumptions of $\lambda_\gamma=+1$ and $K_1(1270)$ dominance, but with considerably less precision. Nevertheless, the hadronic current $\vec{J}$ of $K_1(1270)\to K\pi\pi$ is also obtained in the BESIII study~\cite{BESIII:2025hdt} through an amplitude analysis of $D^{+(0)}\to K^-\pi^+\pi^{0(-)} e^+\nu_e$, which allows a model-dependent estimation of $\mathcal{H}_{\mathrm{ud}}$ with higher precision.

In this work, we present an estimate of the hadronic factor $\mathcal{H}_{\mathrm{ud}}$, based on the hadronic current $\vec{J}$ of $K_1(1270)\to K\pi\pi$ constrained by the charmed-meson semileptonic decays $D^{+(0)}\to K^-\pi^+\pi^{0(-)} e^+\nu_e$. A potential $K_1(1400)$ tail is also investigated by scanning its magnitude and phase relative to $K_1(1270)$.

\section{Hadronic Current Construction}

In the $K_1(1270)$ rest frame, the hadronic current $\vec{J}$ is defined through $\mathcal{A}(K_1(1270)\to K\pi\pi)=\vec{\epsilon}_{K_1}\cdot\vec{J}$. Following Ref.~\cite{BESIII:2025hdt}, the hadronic current is constructed as a coherent sum of quasi-two-body amplitudes $J^\mu=\sum_r g_r J_r^\mu$, where $\vec{J}$ is the spatial part of $J^\mu$ in the $K_1$ rest frame. Each $J_r^\mu$ is built using the covariant tensor formalism~\cite{Zou:2002ar} for $K_1\to Rh_1$, $R\to h_2 h_3$, with detailed expressions given in Table~\ref{tab:K1270_to_Kpipi_amp}. For all two-body decays $a\to bc$, the Blatt-Weisskopf barrier factor $X_L(p)$ is included. The explicit expressions are
\begin{equation}
\begin{split}
X_{L=0}(q) &= 1, \\
X_{L=1}(q) &= \sqrt{\frac{z^2_0+1}{z^2+1}}, \\
X_{L=2}(q) &= \sqrt{\frac{z^4_0+3z^2_0+9}{z^4+3z^2+9}},
\end{split}
\end{equation}
where $z=qR$, $z_0=q_0R$, and the radius parameter $R$ is fixed at $3.0$ GeV$^{-1}$. Here, $q$ is the momentum of $b(c)$ in the $a$ rest frame, and $q_0$ is its value at the nominal mass of $a$.

\begin{table*}[htbp]
	\centering
	\renewcommand{\arraystretch}{1.2}
	\caption{Covariant tensor amplitudes for the processes considered in the decay $K_1(1270)^{-}\to \pi^+_1\pi^-_2K^-_3$.}
	\begin{tabular}{l|c|c}
		\hline
		\hline
		Process  & Formula & Coupling\\
		\hline
		$K_1(1270)^{-}\to \rho^{0}_{12} K^-_3$~($S$-wave), $\rho_{12}\to\pi^+_1 \pi^-_2$  & \multirow{3}{*}{$\sum_{m^\prime} \epsilon(\lambda_{K_1})_{\mu}\epsilon^{\prime *\mu}(m^\prime) \mathcal{R}(M) \epsilon^{\prime *\alpha}(m^\prime)\widetilde{t}_{\alpha}(q^\prime)$} & $ g^{S}_{\rho}$\\
		$K_1(1270)^{-}\to \pi^-_2\bar{K}^{*0}_{13}$~($S$-wave), $\bar{K}^{*0}_{13}\to\pi^+_1 K^-_3$  &  &$g^{S}_{K^*}$\\
		$K_1(1270)^{-}\to \omega^{0}_{12} K^-_3$~($S$-wave), $\omega_{12}\to\pi^+_1 \pi^-_2$ & & $g^{S}_{\omega}$\\
		\hline
		$K_1(1270)^{-}\to \rho^0_{12} K^-_{3}$~($D$-wave), $\rho^0_{12}\to\pi^+_1 \pi^-_2$ & \multirow{2}{*}{  $\sum_{m^\prime}\epsilon(\lambda_{K_1})^{\mu}\epsilon^{\prime *\nu}(m^\prime)\widetilde{t}_{\mu\nu}(q) \mathcal{R}(M) \epsilon^{\prime *\alpha}(m^\prime)\widetilde{t}_{\alpha}(q^\prime)$} & $g^{D}_{\rho}$  \\
		$K_1(1270)^{-}\to \pi^-_2\bar{K}^{*0}_{13}$~($D$-wave), $\bar{K}^{*0}_{13}\to\pi^+_1 K^-_3$  & & $g^{D}_{K^*}$ \\
		\hline
		$K_1(1270)^{-}\to \pi^-_2{(\bar{K}\pi)_ {S-{\rm wave}}}_{13}$, ${(\bar{K}\pi)_ {S-{\rm wave}}}_{13}\to\pi^+_1 K^-_3$ & $\epsilon(\lambda_{K_1})_{\mu}\widetilde{t}^{\mu}(q)\mathcal{R}(M)$ & $ g^{P}_{K_0}$ \\
		\hline
		\hline
	\end{tabular}
	\label{tab:K1270_to_Kpipi_amp}
\end{table*}

Various intermediate states, including $\rho(770)$, $K^*(892)$, $\omega(782)$, and $(K\pi)_{S-{\rm wave}}$, are considered in Ref.~\cite{BESIII:2025hdt}. Their lineshapes $\mathcal{R}(M)$ are described below: $\rho(770)$ is parameterized with a Gounaris-Sakurai model~\cite{Gounaris:1968mw}; $K^*(892)$ is described with a mass-dependent width $\Gamma(m)=\Gamma_0(\frac{q}{q_0})^{2L+1}(\frac{m_0}{m})(\frac{X_L(q)}{X_L(q_0)})^2$; $(K\pi)_{S-{\rm wave}}$ is modeled by a parameterisation from scattering data as $A(m)=F\sin\delta_Fe^{i\delta_F}+R\sin\delta_Re^{i\delta_R}e^{i2\delta_F}$, with parameters fixed to Ref.~\cite{BaBar:2018cka}; and $\omega$ is described by a Breit-Wigner function with constant width. Based on significance tests, the nominal amplitude model contains $K\rho(770)$ in the $S$-wave and $K^*(892)\pi$ in both the $S$-and $D$-waves. Components with statistical significances below 5$\sigma$, including $\pi(K\pi)_{S-{\rm wave}}$, $K\rho(770)$ in the $D$-wave, and $K\omega(782)$ in the $S$-wave, are added to the nominal solution as alternative models to assess the dependence on the amplitude composition. For each model, the complex couplings $g_r$ are determined in amplitude fits to $D^{+(0)}\to K^-\pi^+\pi^{0(-)} e^+\nu_e$.

\section{Calculation of the Hadronic Factor}

Using ${\mathcal{A}}_{ud}$ as defined in Eq.~\ref{eq:Aud_definition}, the averaged up-down asymmetry reported by LHCb~\cite{LHCb:2014vnw} could be written as
\begin{equation}
	\bar{\mathcal{A}}_{ud} = \int{ {\mathcal{A}}_{ud} \times \mathcal{W} d\Phi_{4}}/\int{ \mathcal{W} d\Phi_{4}},
\end{equation}
where $\mathcal{W}=|BW_{K_1(1270)}|^2\times |\mathcal{A}(K_1(1270)\to K\pi\pi)|^2$ is the decay intensity after canceling the constant non-perturbative amplitude of $B\to\gamma K_1(1270)$. The $K_1(1270)$ lineshape $BW_{K_1(1270)}$ is described by a Breit-Wigner function with mass and width fixed to Ref.~\cite{BESIII:2025hdt}. Given $|\mathcal{A}_{K_1(1270)\to K\pi\pi}|^2=|\vec{J}|^2$ in the $K_1(1270)$ rest frame, we obtain $\bar{\mathcal{A}}_{ud} = \frac{3}{4}\lambda_{\gamma}\bar{\mathcal{H}}_{ud}$ with
\begin{equation}
\label{eq:Hud_averaged}
\bar{\mathcal{H}}_{ud}= \frac{\int{{\rm Im}[\hat{n}\cdot(\vec{J}\times\vec{J}^*)]\times |BW_{K_1(1270)}|^2d\Phi_{4}}}{\int{|\vec{J}|^2\times |BW_{K_1(1270)}|^2d\Phi_{4}}}.
\end{equation}
The four-body phase space $d\Phi_{4}(B\to\gamma K\pi\pi)$ can be further factorized as $\frac{d s_{K_1}}{2\pi} d\Phi_2(B\to\gamma K_1) d\Phi_3(K_1\to K\pi\pi)$.

The phase-space integral in Eq.~\ref{eq:Hud_averaged} is evaluated using a Monte Carlo method. One million events of $B\to\gamma K\pi\pi$ are generated uniformly in phase space, with $M_{K\pi\pi}$ in $[1.1,1.3]$~GeV/$c^2$. For each event, the corresponding ${\rm Im}[\hat{n}\cdot(\vec{J}\times\vec{J}^*)]$, $|\vec{J}|^2$, and $|BW_{K_1}|^2$ are calculated. The convention $\hat{n}=\hat{\vec{p}}_{\pi,{\rm slow}}\times\hat{\vec{p}}_{\pi,{\rm fast}}$ is used for consistency with Ref.~\cite{LHCb:2014vnw}. By summing over all events, the averaged $\bar{\mathcal{H}}_{ud}$ could be obtained. The corresponding statistical uncertainty is estimated by performing 1000 alternative calculations, with the complex couplings $g_{r}$ and the $K_1$ lineshape varied according to the covariance matrix. The standard deviation of the $\bar{\mathcal{H}}_{ud}$ distribution from these alternative calculations is taken as the statistical uncertainty.

\section{Numerical Results}

The fitted complex couplings and resonance parameters of $K_1(1270)$ for the nominal and alternative models in Ref.~\cite{BESIII:2025hdt} are summarized in Table~\ref{tab:model_pars}, with correlation matrices given in Appendix~\ref{appendix}. The calculated $\bar{\mathcal{H}}_{ud}$ values for the various amplitude models are also summarized in Table~\ref{tab:model_pars}. Although significant model dependence is seen among these models, all the estimates are consistent with the model-independent extraction $(0.01\pm0.11)$ in $D\to\bar{K}_1(1270)e^+\nu_e$~\cite{BESIII:2025hdt} and have much higher precision. They are also systematically lower than the expectation $(0.092\pm0.023)$ obtained by assuming $\lambda_{\gamma}=1$ in the SM.

\begin{table}[htbp]
	\centering
	\fontsize{8}{10}\selectfont
	\renewcommand{\arraystretch}{1.2}
	\caption{Complex couplings and resonance parameters of $K_1(1270)$ of the nominal and alternative models.}
	\label{tab:model_pars}
	\begin{tabular}{l|r@{$\,\pm\,$}lr@{$\,\pm\,$}lr@{$\,\pm\,$}lr@{$\,\pm\,$}l}
		\hline\hline
		Par. & \multicolumn{2}{c}{Nominal} & \multicolumn{2}{c}{Alter I} & \multicolumn{2}{c}{Alter II} & \multicolumn{2}{c}{Alter III} \\
		\hline
		$|g^{S}_{\rho}|$ & \multicolumn{8}{c}{Fixed to 1} \\
		$\angle g^{S}_{\rho}$ & \multicolumn{8}{c}{Fixed to 0} \\
		\hline
		$|g^{D}_{\rho}|$ & \multicolumn{2}{c}{---} & $0.48$ & $0.25$ & \multicolumn{2}{c}{---} & \multicolumn{2}{c}{---} \\
		$\angle g^{D}_{\rho}$ & \multicolumn{2}{c}{---} & $-0.81$ & $0.40$ & \multicolumn{2}{c}{---} & \multicolumn{2}{c}{---} \\
		\hline
		$|g^{S}_{K^*}|$ & $0.19$ & $0.02$ & $0.17$ & $0.02$ & $0.30$ & $0.04$ & $0.19$ & $0.02$ \\
		$\angle g^{S}_{K^*}$ & $0.45$ & $0.11$ & $0.48$ & $0.11$ & $0.79$ & $0.15$ & $0.49$ & $0.11$ \\
		\hline
		$|g^{D}_{K^*}|$ & $1.01$ & $0.13$ & $1.01$ & $0.11$ & $1.03$ & $0.14$ & $1.00$ & $0.14$ \\
		$\angle g^{D}_{K^*}$ & $-2.07$ & $0.11$ & $-2.12$ & $0.12$ & $-1.82$ & $0.18$ & $-2.02$ & $0.11$ \\
		\hline
		$|g^{P}_{K_0}|$ & \multicolumn{2}{c}{---} & \multicolumn{2}{c}{---} & $2.17$ & $0.91$ & \multicolumn{2}{c}{---} \\
		$\angle g^{P}_{K_0}$ & \multicolumn{2}{c}{---} & \multicolumn{2}{c}{---} & $0.26$ & $0.22$ & \multicolumn{2}{c}{---} \\
		\hline
		$|g^{S}_{\omega}|$ & \multicolumn{2}{c}{---} & \multicolumn{2}{c}{---} & \multicolumn{2}{c}{---} & $0.04$ & $0.02$ \\
		$\angle g^{S}_{\omega}$ & \multicolumn{2}{c}{---} & \multicolumn{2}{c}{---} & \multicolumn{2}{c}{---} & $2.07$ & $0.43$ \\
		\hline
		$m_{K_1}$ & $1.271$ & $0.003$ & $1.271$ & $0.003$ & $1.276$ & $0.004$ & $1.270$ & $0.003$ \\
		$\Gamma_{K_1}$ & $0.168$ & $0.010$ & $0.170$ & $0.010$ & $0.150$ & $0.010$ & $0.170$ & $0.010$ \\
		\hline
		$\bar{\mathcal{H}}_{ud}$ & $-0.033$ & $0.006$ & $0.008$ & $0.031$ & $0.017$ & $ 0.024$ &  $-0.033$ & $0.006$ \\
		\hline\hline
	\end{tabular}
\end{table}

\begin{figure*}[htbp]
	\centering
	\centering
	\setlength{\abovecaptionskip}{-0.0cm}
	\includegraphics[width=13.0cm]{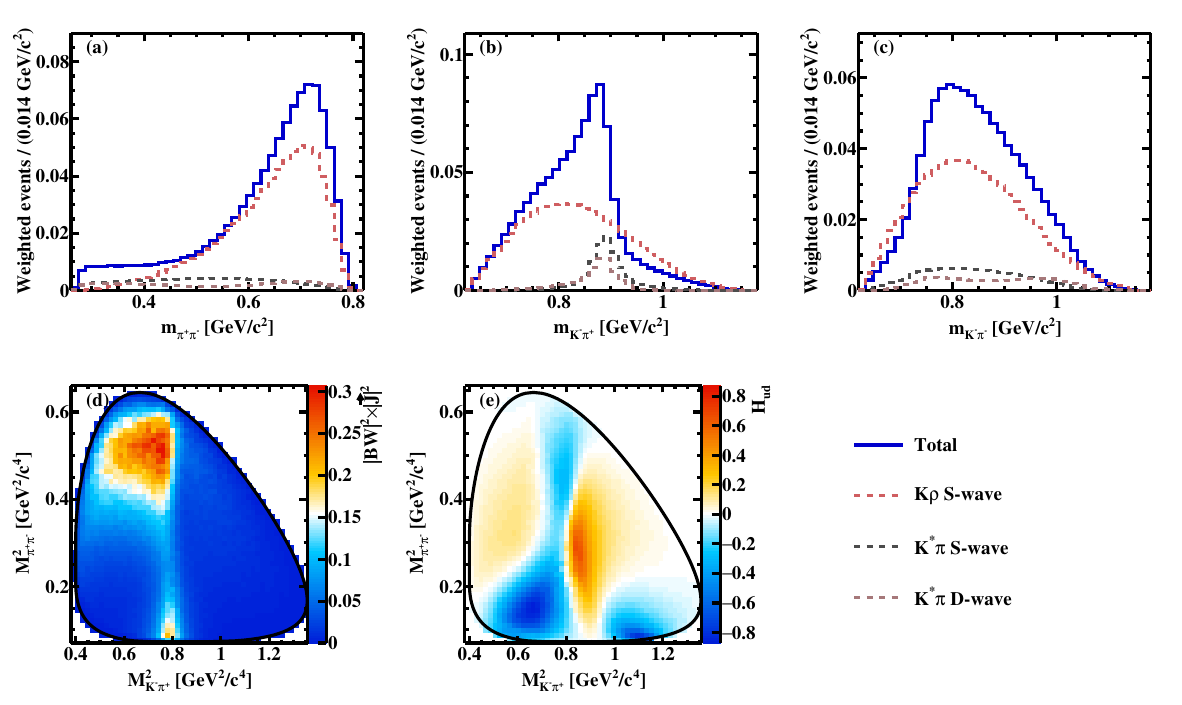}
	\caption{(a-c) Two-body invariant-mass spectra, (d) Dalitz plot, and (e) two-dimensional $\mathcal{H}_{ud}$ distribution of $K_1(1270)^-\to K^-\pi^+\pi^-$ based on the nominal model.}
	\label{fig:Mass_Dalitz_023}
\end{figure*}

\begin{figure*}[htbp]
	\centering
	\centering
	\setlength{\abovecaptionskip}{-0.0cm}
	\includegraphics[width=13.0cm]{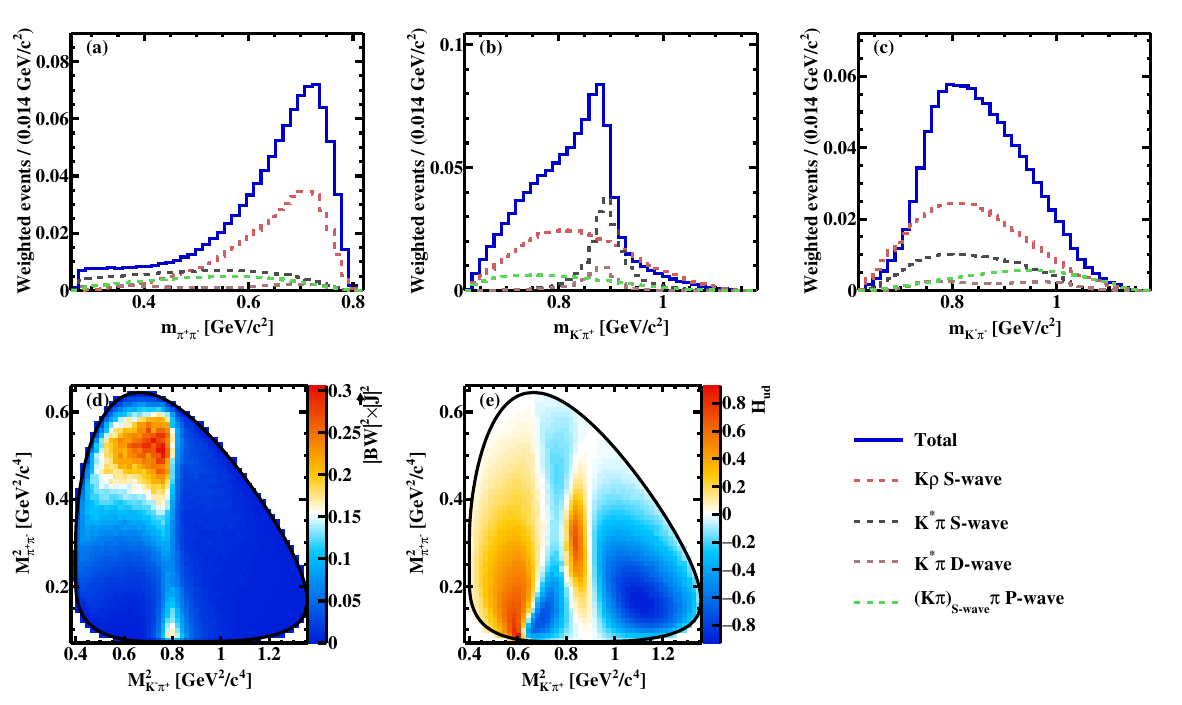}
	\caption{The same as Fig.~\ref{fig:Mass_Dalitz_023}, but for alternative model II with an additional $(K\pi)_{\rm S-wave}$ component.}
	\label{fig:Mass_Dalitz_0234}
\end{figure*}

To illustrate the model dependence, we compare the nominal model with alternative model II, which produces the largest shift in $\bar{\mathcal{H}}_{\mathrm{ud}}$. As shown in
Figs.~\ref{fig:Mass_Dalitz_023} and~\ref{fig:Mass_Dalitz_0234}, the two-body invariant-mass spectra and Dalitz-plot intensities predicted by the two models are similar, as expected since both are constrained by the same data sample. However, their hadronic factor distributions $\mathcal{H}_{\mathrm{ud}}(M^2_{\pi^+\pi^-},M^2_{K^-\pi^+})$ differ significantly, particularly at low $\pi^+\pi^-$ mass. In the present amplitude decomposition, an individual component does not contribute to $\mathcal{H}_{ud}\propto {\rm Im}[\hat{n}\cdot(\vec{J}\times\vec{J}^*)]$. Nonzero $\mathcal{H}_{ud}$ arises from interference between different components. While the overall intensity remains similar, the additional $(K\pi)_{S-{\rm wave}}$ component produces a different interference pattern in alternative model II and hence changes the local $\mathcal{H}_{ud}$ significantly. We further decompose the averaged factor $\bar{\mathcal{H}}_{\mathrm{ud}}$ into the interference contributions between intermediate states $r_1$ and $r_2$ as
\begin{equation}
\label{eq:Hud_averaged_interference}
\bar{\mathcal{H}}^{r_1r_2}_{ud}= \frac{\int{{\rm Im}[\hat{n}\cdot(\vec{J_{r_1}}\times\vec{J}^*_{r_2}+\vec{J_{r_2}}\times\vec{J}^*_{r_1})]\times |BW_{K_1}|^2d\Phi_{4}}}{\int{|\vec{J}|^2\times |BW_{K_1}|^2d\Phi_{4}}},
\end{equation}
The values are summarized in Table~\ref{tab:check_interference}.

\begin{table}[htbp]
	\centering
	\renewcommand{\arraystretch}{1.2}
	\setlength{\tabcolsep}{5pt}
	\caption{Interference contributions $\bar{\mathcal{H}}^{r_1r_2}_{ud}$ between intermediate states $r_1$ and $r_2$ in the nominal and alternative models. $R_{\mathrm{add}}$ corresponds to $\rho_D$, $K_{0,P}$, and $\omega_S$ in alternative models I, II, and III, respectively.}
	\label{tab:check_interference}
	\begin{tabular}{c|cccc}
		\hline\hline
		$r_1 \times r_2$ & Nominal & Alter I & Alter II & Alter III \\
		\hline
		$\rho_S \times K^*_S$ & $\phantom{-}0.0290$ & $\phantom{-}0.0255$ & $-0.0021$ & $\phantom{-}0.0258$ \\
		$\rho_S \times K^*_D$ & $-0.0184$ & $-0.0230$ & $\phantom{-}0.0007$ & $-0.0144$ \\
		$K^*_S \times K^*_D$ & $-0.0438$ & $-0.0367$ & $-0.0419$ & $-0.0438$ \\
		\hline
		$\rho_S \times R_{\mathrm{add}}$ & --- & $\phantom{-}0.0458$ & $-0.0022$ & $\phantom{-}0.0000$ \\
		$K^*_S \times R_{\mathrm{add}}$ & --- & $-0.0011$ & $\phantom{-}0.0780$ & $\phantom{-}0.0032$ \\
		$K^*_D \times R_{\mathrm{add}}$ & --- & $-0.0023$ & $-0.0154$ & $-0.0033$ \\
		\hline
		Sum & $-0.0332$ & $\phantom{-}0.0083$ & $\phantom{-}0.0171$ & $-0.0326$ \\
		\hline\hline
	\end{tabular}
\end{table}

Nevertheless, the averaged factor $\bar{\mathcal{H}}_{\mathrm{ud}}$ is the intensity-weighted average of the local $\mathcal{H}_{\mathrm{ud}}$ over the Dalitz plot. The largest differences between the models occur mainly in low-intensity regions and therefore have a reduced impact on the integrated value. Although the local $\mathcal{H}_{\mathrm{ud}}$ differs by more than 0.5 between models in some regions of the Dalitz plot, the largest shift in its intensity-weighted average, $\bar{\mathcal{H}}_{\mathrm{ud}}$, is approximately 0.05, enabling reliable estimates using a model-dependent approach. For reference, the distributions of alternative models I and III are given in Appendix~\ref{appendix}.

\section{Scan of the $K_1(1400)$ Contribution}

Given that an isolated $K_1(1270)$ can hardly explain the observed $\bar{\mathcal{A}}_{ud}=(0.069\pm0.017)$ in the mass region $M_{K^+\pi^+\pi^-}\in[1.1,1.3]$~GeV/$c^2$, potential tails from other excited kaon states are further investigated. According to an amplitude analysis of $B^+\to K^+\pi^+\pi^- J/\psi$~\cite{Belle:2010wrf}, whose $M_{K^+\pi^+\pi^-}$ spectrum is similar to that of $B^+\to K^+\pi^+\pi^- \gamma$, $K_1(1400)$ provides the largest contribution apart from $K_1(1270)$, with a considerable fit fraction of up to 30\%.

\begin{figure}[htbp]
	\centering
	\centering
	\setlength{\abovecaptionskip}{-0.0cm}
	\includegraphics[width=0.49\linewidth]{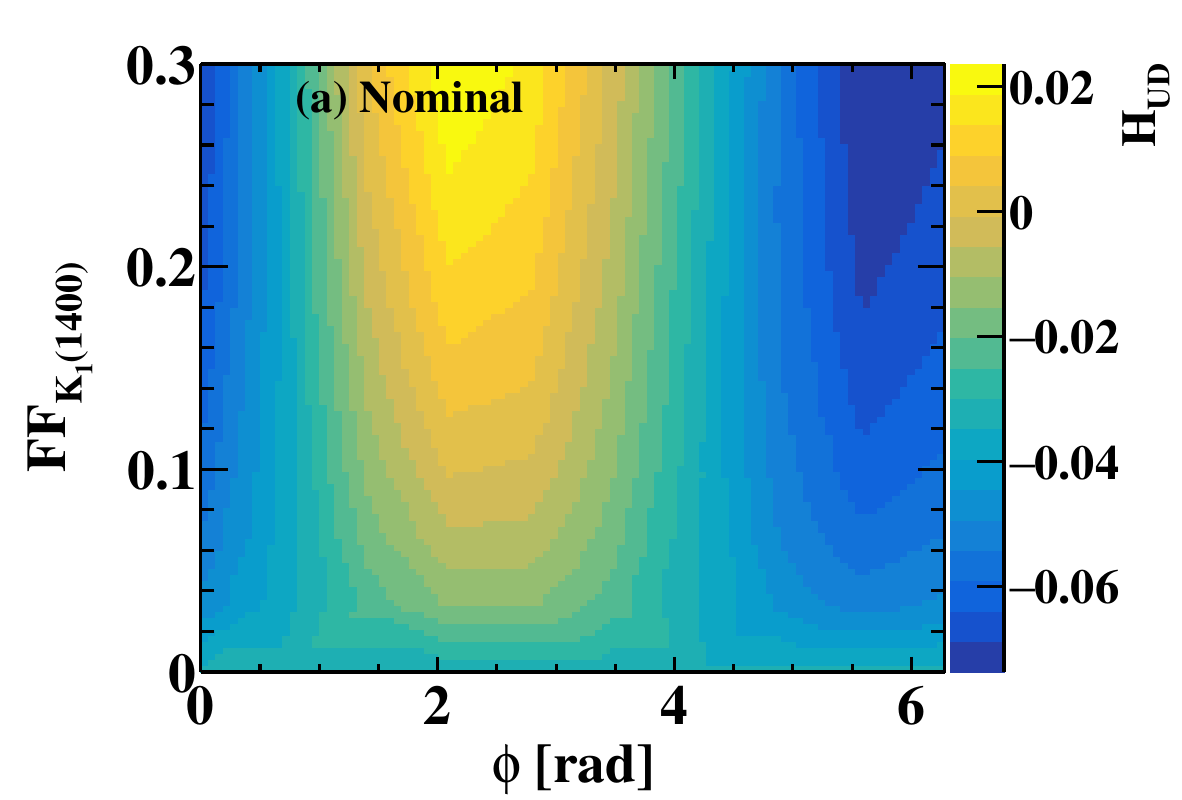}
	\includegraphics[width=0.49\linewidth]{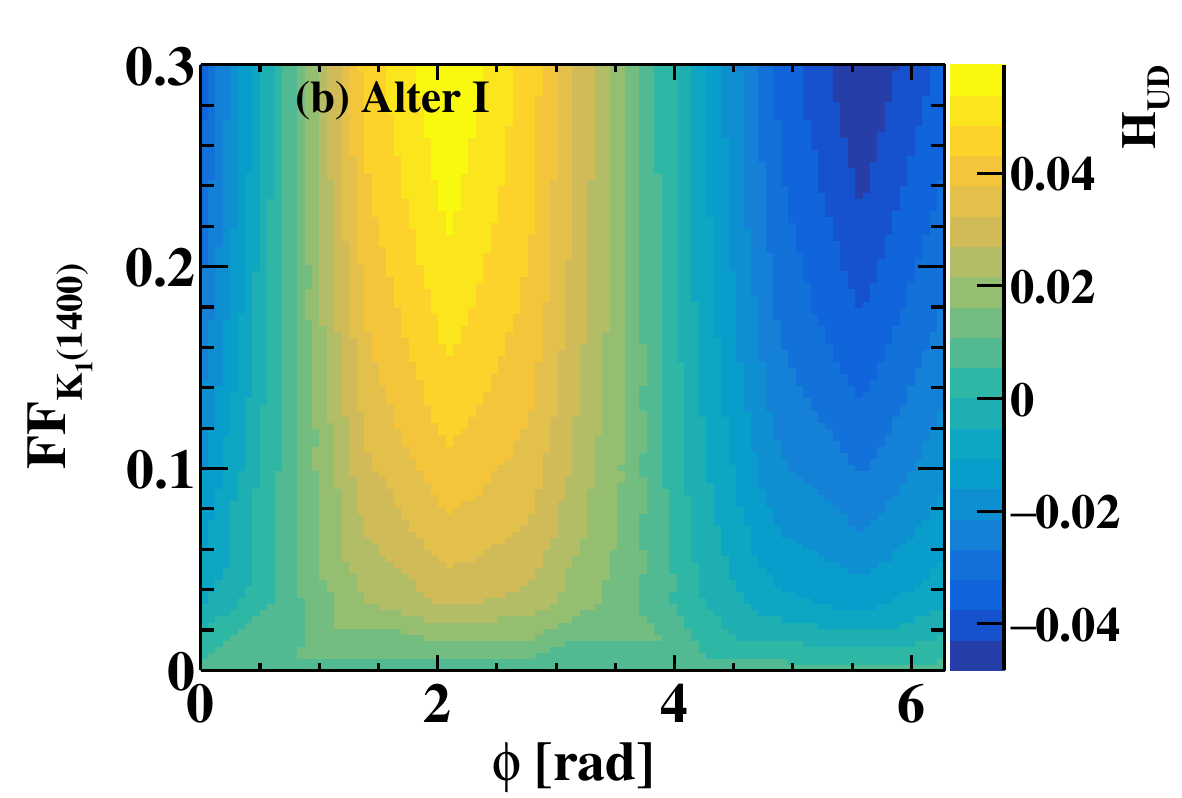} \\
	\includegraphics[width=0.49\linewidth]{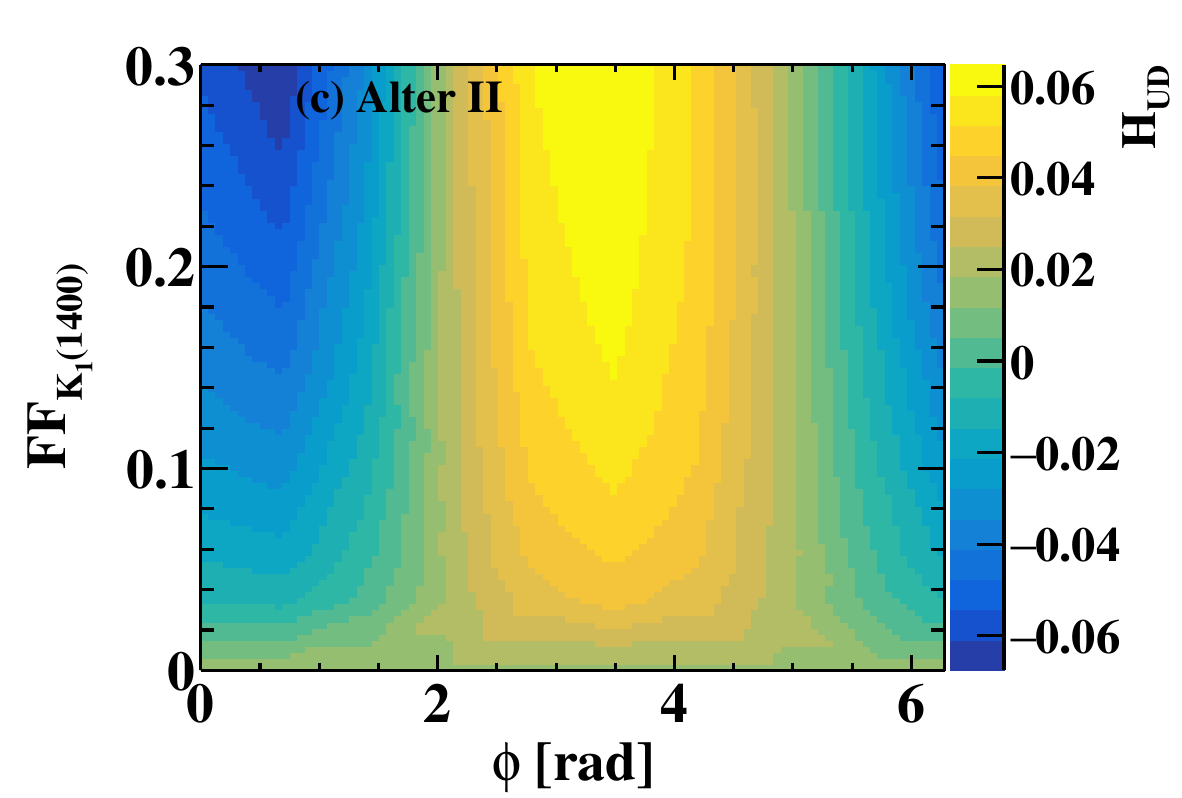}
	\includegraphics[width=0.49\linewidth]{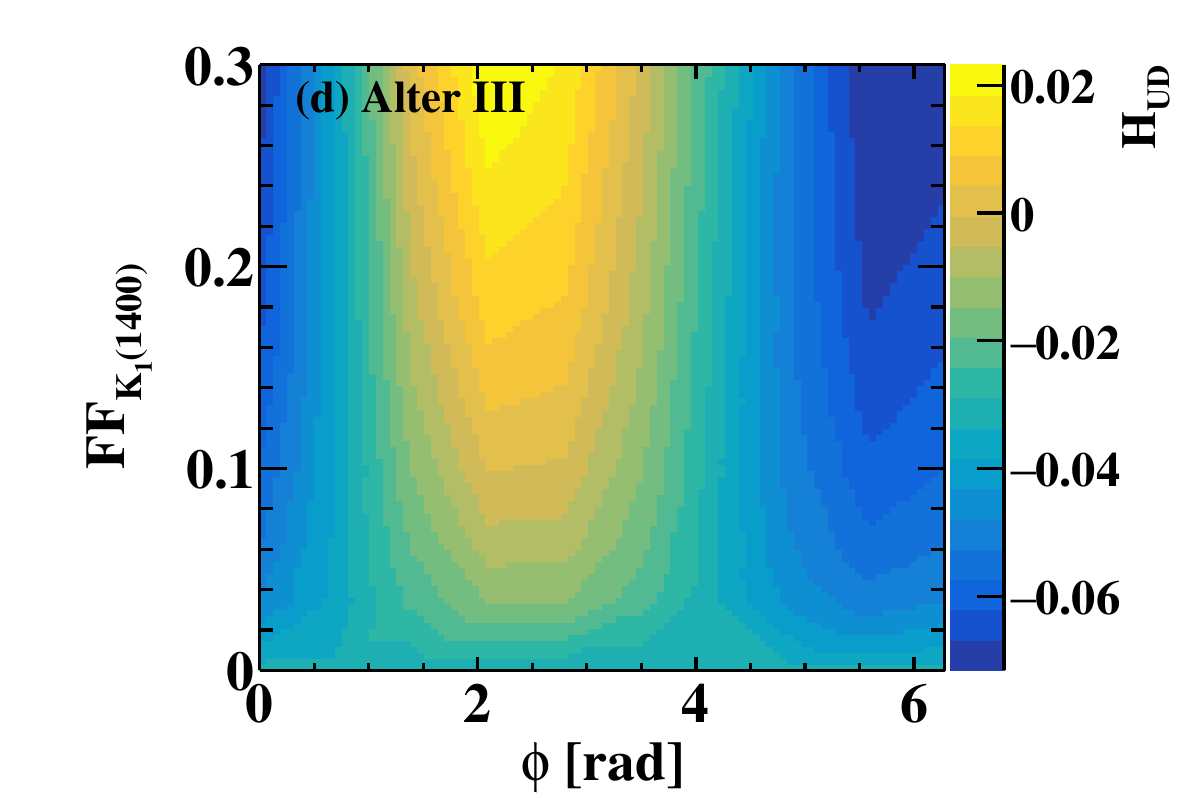}
	\caption{Two-dimensional scans over the relative phase $\phi$ and the fit fraction ${\rm FF}_{K_1(1400)}$ based on (a) the nominal model and alternative models (b) I, (c) II, and (d) III.}
	\label{fig:scan_K1400}
\end{figure}

To estimate the impact of $K_1(1400)$ on $\bar{\mathcal{H}}_{\mathrm{ud}}$, the hadronic current $\vec{J}_{\rm tot}$ is further revised as
\begin{equation}
	BW_{K_1(1270)}\times\vec{J}_{K_1(1270)}+\rho e^{i\phi}BW_{K_1(1400)}\times\vec{J}_{K_1(1400)},
\end{equation}
where $\rho$ and $\phi$ are the magnitude and phase of $K_1(1400)$ relative to $K_1(1270)$. The hadronic current $\vec{J}_{K_1(1400)}$ includes its dominant decay mode $K^*(892)\pi$ via the $S$-wave~\cite{ParticleDataGroup:2026mpi}, which has the same form as that of $K_1(1270)$. Hence, the averaged $\bar{\mathcal{H}}_{ud}$ could be calculated as
\begin{equation}
\label{eq:Hud_averaged_scan}
\bar{\mathcal{H}}_{ud}= \frac{\int{{\rm Im}[\hat{n}\cdot(\vec{J}_{\rm tot}\times\vec{J}_{\rm tot}^*)]d\Phi_{4}}}{\int{|\vec{J}_{\rm tot}|^2d\Phi_{4}}}.
\end{equation}

A two-dimensional scan is performed over the relative phase $\phi$ and the fit fraction ${\rm FF}_{K_1(1400)}$, from which $\rho$ could be uniquely determined. Figure~\ref{fig:scan_K1400} shows the scan results based on both the nominal and alternative models. Introducing a $K_1(1400)$ contribution could significantly reduce the tension between the estimated and expected $\bar{\mathcal{H}}_{\mathrm{ud}}$, especially for alternative models I and II, which yield values of $\bar{\mathcal{H}}_{\mathrm{ud}}$ up to 0.06 within ${\rm FF}_{K_1(1400)}<30\%$.

\section{Summary and Discussion}

In summary, using $K_1(1270)\to K\pi\pi$ amplitude models constrained by the charmed-meson semileptonic decays $D^{+(0)}\to K^-\pi^+\pi^{0(-)} e^+\nu_e$, we present a model-dependent estimate of the up-down asymmetry observed in $B^+\to K^+\pi^+\pi^-\gamma$ in the mass region $M_{K^+\pi^+\pi^-}\in[1.1,1.3]$~GeV/$c^2$. With an isolated $K_1(1270)$ contribution, estimates based on various amplitude models suggest significant model dependence of $\mathcal{H}_{\rm ud}$ in some Dalitz-plot regions, but the intensity-averaged $\bar{\mathcal{H}}_{\rm ud}$ remains stable compared to the local variations. The obtained $\bar{\mathcal{H}}_{\rm ud}$ values are consistent with the model-independent extraction $(0.01\pm0.11)$ in $D\to\bar{K}_1(1270)e^+\nu_e$~\cite{BESIII:2025hdt} and have much higher precision, but are systematically lower than the expectation $(0.092\pm0.023)$ obtained by assuming $\lambda_{\gamma}=1$ in the SM. A further two-dimensional scan is performed over the phase angle and fit fraction of a potential $K_1(1400)$ component; the tension could be significantly reduced with a reasonable fit fraction of $K_1(1400)$.

Our result disfavors an isolated $K_1(1270)$ contribution in the mass region $M_{K^+\pi^+\pi^-}\in[1.1,1.3]$~GeV/$c^2$ and suggests that a sizable $K_1(1400)$ tail should be considered. Following this conclusion, the model-independent extraction proposed in Refs.~\cite{Bian:2021gwf,Wang:2019wee,Kou:2016iau} should be further corrected by considering other $K^*$ contributions to access a reliable photon-polarization in $B^+\to K^+\pi^+\pi^-\gamma$.

\newpage

\bibliography{reference}

\onecolumngrid
\appendix
\section*{Appendix}
\label{appendix}

\begin{figure*}[htbp]
	\centering
	\centering
	\setlength{\abovecaptionskip}{-0.0cm}
	\includegraphics[width=13.0cm]{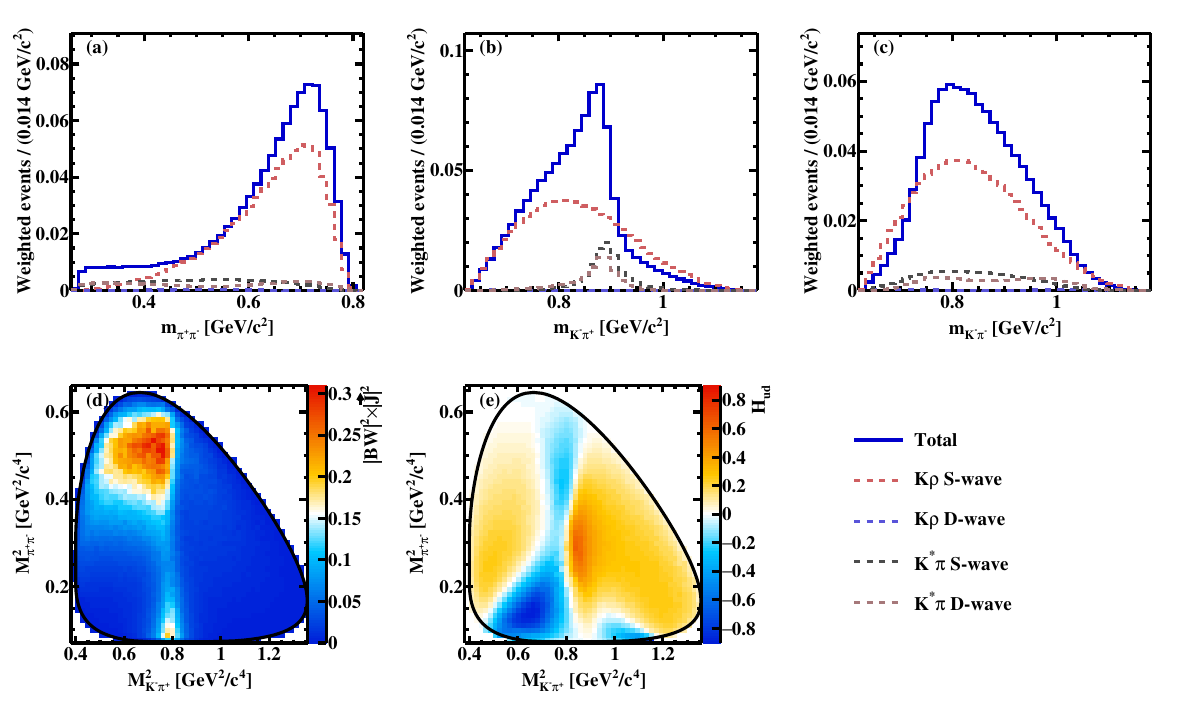}
	\caption{The same as Fig.~\ref{fig:Mass_Dalitz_023}, but for alternative model I with an additional $K\rho$ $D$-wave component.}
	\label{fig:Mass_Dalitz_0123}
\end{figure*}

\begin{figure*}[htbp]
	\centering
	\centering
	\setlength{\abovecaptionskip}{-0.0cm}
	\includegraphics[width=13.0cm]{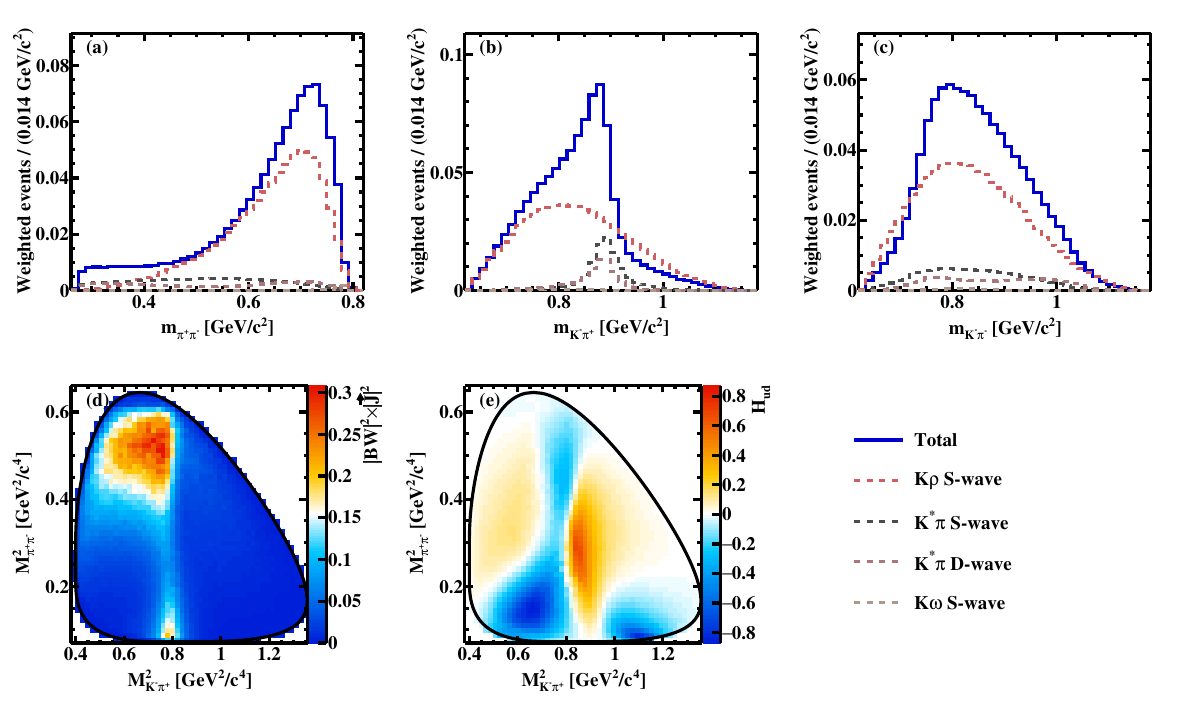}
	\caption{The same as Fig.~\ref{fig:Mass_Dalitz_023}, but for alternative model III with an additional $K\omega$ $S$-wave component.}
	\label{fig:Mass_Dalitz_0235}
\end{figure*}

Correlation matrices for both the nominal and alternative models are given in Tables~\ref{tab:corr_023},~\ref{tab:corr_0123},~\ref{tab:corr_0234}, and~\ref{tab:corr_0235}. Figures~\ref{fig:Mass_Dalitz_0123} and~\ref{fig:Mass_Dalitz_0235} show the two-body invariant-mass spectra, Dalitz-plot intensities, and hadronic factor distributions $\mathcal{H}_{\mathrm{ud}}(M^2_{\pi^+\pi^-},M^2_{K^-\pi^+})$ for alternative models I and III, respectively.

\begin{table}[htbp]
	\centering
	\fontsize{8}{10}\selectfont
	\renewcommand{\arraystretch}{1.2}
	\caption{Correlation matrix of the nominal model.}
	\label{tab:corr_023}
	\begin{tabular}{l|cccccc}
		\hline\hline
		Par. & $|g^{S}_{K^*}|$ & $\angle g^{S}_{K^*}$ & $|g^{D}_{K^*}|$ & $\angle g^{D}_{K^*}$ & $m_{K_1}$ & $\Gamma_{K_1}$ \\
		\hline
		$|g^{S}_{K^*}|$ & $\phantom{-}1.00$ & $\phantom{-}0.35$ & $-0.54$ & $-0.01$ & $\phantom{-}0.27$ & $-0.31$ \\
		$\angle g^{S}_{K^*}$ & $\phantom{-}0.35$ & $\phantom{-}1.00$ & $-0.45$ & $\phantom{-}0.30$ & $-0.08$ & $-0.08$ \\
		$|g^{D}_{K^*}|$ & $-0.54$ & $-0.45$ & $\phantom{-}1.00$ & $\phantom{-}0.05$ & $-0.16$ & $\phantom{-}0.23$ \\
		$\angle g^{D}_{K^*}$ & $-0.01$ & $\phantom{-}0.30$ & $\phantom{-}0.05$ & $\phantom{-}1.00$ & $-0.13$ & $\phantom{-}0.02$ \\
		$m_{K_1}$ & $\phantom{-}0.27$ & $-0.08$ & $-0.16$ & $-0.13$ & $\phantom{-}1.00$ & $\phantom{-}0.04$ \\
		$\Gamma_{K_1}$ & $-0.31$ & $-0.08$ & $\phantom{-}0.23$ & $\phantom{-}0.02$ & $\phantom{-}0.04$ & $\phantom{-}1.00$ \\
		\hline\hline
	\end{tabular}
\end{table}

\begin{table}[htbp]
	\centering
	\fontsize{8}{10}\selectfont
	\renewcommand{\arraystretch}{1.2}
	\caption{Correlation matrix of alternative model I.}
	\label{tab:corr_0123}
	\begin{tabular}{l|cccccccc}
		\hline\hline
		Par. & $|g^{D}_{\rho}|$ & $\angle g^{D}_{\rho}$ & $|g^{S}_{K^*}|$ & $\angle g^{S}_{K^*}$ & $|g^{D}_{K^*}|$ & $\angle g^{D}_{K^*}$ & $m_{K_1}$ & $\Gamma_{K_1}$ \\
		\hline
		$|g^{D}_{\rho}|$ & $\phantom{-}1.00$ & $-0.56$ & $-0.07$ & $\phantom{-}0.04$ & $\phantom{-}0.06$ & $-0.30$ & $\phantom{-}0.03$ & $-0.01$ \\
		$\angle g^{D}_{\rho}$ & $-0.56$ & $\phantom{-}1.00$ & $-0.28$ & $-0.10$ & $\phantom{-}0.03$ & $\phantom{-}0.07$ & $-0.07$ & $\phantom{-}0.11$ \\
		$|g^{S}_{K^*}|$ & $-0.07$ & $-0.28$ & $\phantom{-}1.00$ & $\phantom{-}0.19$ & $-0.42$ & $\phantom{-}0.15$ & $\phantom{-}0.24$ & $-0.30$ \\
		$\angle g^{S}_{K^*}$ & $\phantom{-}0.04$ & $-0.10$ & $\phantom{-}0.19$ & $\phantom{-}1.00$ & $-0.33$ & $\phantom{-}0.37$ & $-0.16$ & $-0.03$ \\
		$|g^{D}_{K^*}|$ & $\phantom{-}0.06$ & $\phantom{-}0.03$ & $-0.42$ & $-0.33$ & $\phantom{-}1.00$ & $-0.10$ & $-0.11$ & $\phantom{-}0.19$ \\
		$\angle g^{D}_{K^*}$ & $-0.30$ & $\phantom{-}0.07$ & $\phantom{-}0.15$ & $\phantom{-}0.37$ & $-0.10$ & $\phantom{-}1.00$ & $-0.12$ & $-0.03$ \\
		$m_{K_1}$ & $\phantom{-}0.03$ & $-0.07$ & $\phantom{-}0.24$ & $-0.16$ & $-0.11$ & $-0.12$ & $\phantom{-}1.00$ & $\phantom{-}0.05$ \\
		$\Gamma_{K_1}$ & $-0.01$ & $\phantom{-}0.11$ & $-0.30$ & $-0.03$ & $\phantom{-}0.19$ & $-0.03$ & $\phantom{-}0.05$ & $\phantom{-}1.00$ \\
		\hline\hline
	\end{tabular}
\end{table}

\begin{table}[htbp]
	\centering
	\fontsize{8}{10}\selectfont
	\renewcommand{\arraystretch}{1.2}
	\caption{Correlation matrix of alternative model II.}
	\label{tab:corr_0234}
	\begin{tabular}{l|cccccccc}
		\hline\hline
		Par. & $|g^{S}_{K^*}|$ & $\angle g^{S}_{K^*}$ & $|g^{D}_{K^*}|$ & $\angle g^{D}_{K^*}$ & $|g^{P}_{K_0}|$ & $\angle g^{P}_{K_0}$ & $m_{K_1}$ & $\Gamma_{K_1}$ \\
		\hline
		$|g^{S}_{K^*}|$ & $\phantom{-}1.00$ & $\phantom{-}0.61$ & $-0.16$ & $\phantom{-}0.05$ & $\phantom{-}0.79$ & $\phantom{-}0.39$ & $\phantom{-}0.40$ & $-0.58$ \\
		$\angle g^{S}_{K^*}$ & $\phantom{-}0.61$ & $\phantom{-}1.00$ & $-0.31$ & $\phantom{-}0.44$ & $\phantom{-}0.75$ & $\phantom{-}0.71$ & $-0.03$ & $-0.44$ \\
		$|g^{D}_{K^*}|$ & $-0.16$ & $-0.31$ & $\phantom{-}1.00$ & $\phantom{-}0.03$ & $-0.04$ & $-0.19$ & $\phantom{-}0.04$ & $\phantom{-}0.12$ \\
		$\angle g^{D}_{K^*}$ & $\phantom{-}0.05$ & $\phantom{-}0.44$ & $\phantom{-}0.03$ & $\phantom{-}1.00$ & $\phantom{-}0.30$ & $\phantom{-}0.59$ & $-0.28$ & $-0.10$ \\
		$|g^{P}_{K_0}|$ & $\phantom{-}0.79$ & $\phantom{-}0.75$ & $-0.04$ & $\phantom{-}0.30$ & $\phantom{-}1.00$ & $\phantom{-}0.73$ & $\phantom{-}0.20$ & $-0.57$ \\
		$\angle g^{P}_{K_0}$ & $\phantom{-}0.39$ & $\phantom{-}0.71$ & $-0.19$ & $\phantom{-}0.59$ & $\phantom{-}0.73$ & $\phantom{-}1.00$ & $-0.20$ & $-0.34$ \\
		$m_{K_1}$ & $\phantom{-}0.40$ & $-0.03$ & $\phantom{-}0.04$ & $-0.28$ & $\phantom{-}0.20$ & $-0.20$ & $\phantom{-}1.00$ & $-0.05$ \\
		$\Gamma_{K_1}$ & $-0.58$ & $-0.44$ & $\phantom{-}0.12$ & $-0.10$ & $-0.57$ & $-0.34$ & $-0.05$ & $\phantom{-}1.00$ \\
		\hline\hline
	\end{tabular}
\end{table}

\begin{table}[htbp]
	\centering
	\fontsize{8}{10}\selectfont
	\renewcommand{\arraystretch}{1.2}
	\caption{Correlation matrix of alternative model III.}
	\label{tab:corr_0235}
	\begin{tabular}{l|cccccccc}
		\hline\hline
		Par. & $|g^{S}_{K^*}|$ & $\angle g^{S}_{K^*}$ & $|g^{D}_{K^*}|$ & $\angle g^{D}_{K^*}$ & $|g^{S}_{\omega}|$ & $\angle g^{S}_{\omega}$ & $m_{K_1}$ & $\Gamma_{K_1}$ \\
		\hline
		$|g^{S}_{K^*}|$ & $\phantom{-}1.00$ & $\phantom{-}0.35$ & $-0.55$ & $-0.04$ & $-0.02$ & $-0.09$ & $\phantom{-}0.26$ & $-0.31$ \\
		$\angle g^{S}_{K^*}$ & $\phantom{-}0.35$ & $\phantom{-}1.00$ & $-0.47$ & $\phantom{-}0.30$ & $\phantom{-}0.15$ & $\phantom{-}0.08$ & $-0.08$ & $-0.08$ \\
		$|g^{D}_{K^*}|$ & $-0.55$ & $-0.47$ & $\phantom{-}1.00$ & $\phantom{-}0.07$ & $-0.01$ & $-0.05$ & $-0.17$ & $\phantom{-}0.23$ \\
		$\angle g^{D}_{K^*}$ & $-0.04$ & $\phantom{-}0.30$ & $\phantom{-}0.07$ & $\phantom{-}1.00$ & $\phantom{-}0.22$ & $\phantom{-}0.10$ & $-0.15$ & $\phantom{-}0.04$ \\
		$|g^{S}_{\omega}|$ & $-0.02$ & $\phantom{-}0.15$ & $-0.01$ & $\phantom{-}0.22$ & $\phantom{-}1.00$ & $\phantom{-}0.30$ & $-0.09$ & $\phantom{-}0.06$ \\
		$\angle g^{S}_{\omega}$ & $-0.09$ & $\phantom{-}0.08$ & $-0.05$ & $\phantom{-}0.10$ & $\phantom{-}0.30$ & $\phantom{-}1.00$ & $\phantom{-}0.09$ & $\phantom{-}0.03$ \\
		$m_{K_1}$ & $\phantom{-}0.26$ & $-0.08$ & $-0.17$ & $-0.15$ & $-0.09$ & $\phantom{-}0.09$ & $\phantom{-}1.00$ & $\phantom{-}0.03$ \\
		$\Gamma_{K_1}$ & $-0.31$ & $-0.08$ & $\phantom{-}0.23$ & $\phantom{-}0.04$ & $\phantom{-}0.06$ & $\phantom{-}0.03$ & $\phantom{-}0.03$ & $\phantom{-}1.00$ \\
		\hline\hline
	\end{tabular}
\end{table}

\end{document}